\documentclass{article} 

\usepackage{hyperref}
\usepackage[super,sort,compress]{cite}
\usepackage{subfigure}
\usepackage{color}
\usepackage{soul}
\usepackage{amsmath}
\usepackage{amssymb}
\usepackage{booktabs}
\usepackage{graphicx}   
\usepackage{bm}
\usepackage{color}
\usepackage{cancel}
\usepackage{slashed}
\usepackage{subfigure}
\usepackage{soul}

\newcommand{\Ket}[1]{\left\vert #1\right\rangle}
\newcommand{\Bra}[1]{\left\langle #1\right\vert}

\newcommand{\BraKet}[2]{\left\langle#1\vert #2\right\rangle}
\newcommand{\KetBra}[2]{\left\vert#1\right\rangle\left\langle#2\right\vert}

\newcommand{\ii}{\mathrm{i}}
\newcommand{\ee}{\mathrm{e}}

\newcommand{\ignore}[1]{}

\renewcommand{\title}[1]{{ \Large\bf \begin{center} #1 \end{center}}}
\newcommand{\authors}[1]{{ \begin{center} #1 \end{center}}}
\newcommand{\address}[1]{{ \it \begin{center} #1 \end{center}}}

\usepackage{geometry}
\begin{document}

\title{Evanescent Wave Approximation for Non-Hermitian Hamiltonians}

\authors{Benedetto Militello and  Anna Napoli}

\address{%
Universit\'a degli Studi di Palermo, Dipartimento di Fisica e Chimica - Emilio Segr\`e, Via Archirafi 36, 90123 Palermo\\
INFN Sezione di Catania, Via Santa Sofia 64, 95123 Catania, Italy
}

\abstract{
The counterpart of the rotating wave approximation for non-Hermitian Hamiltonians is considered, which allows for the derivation of a suitable effective Hamiltonian for systems with some states undergoing decays. In the limit of very high decay rates, on the basis of this effective description we can predict the occurrence of a quantum Zeno dynamics which is interpreted as the removal of some coupling terms and the vanishing of an operatorial pseudo-Lamb shift.
}

\section{Introduction}

In the study of the dynamical behaviour of an assigned system, one very often comes across the impossibility of handling, both analytically and numerically, its Hamiltonian model. The difficulties can originate from the complexity of the system stemming from  the presence, for example, of an exceptionally large number of degrees of freedom, time-dependent driving terms, etc. Generally speaking, however, the description of very simple physical systems also leads to the presence of terms in the Hamiltonian that are difficult to deal with.  It is well known that even analyzing the archetypal system in the context of matter-radiation interaction, consisting of a single atom interacting with a single cavity mode, one faces with an Hamiltonian model which is hard to solve~\cite{ref:RabiReview}, unless suitable approximations are made to obtain a solvable effective model, for example the very famous Jaynes-Cummings model~\cite{ref:Knight1993}. 
On the other hand, if one wants to catch some properties of a system or some aspects of its dynamical behaviour, it is often not necessary to consider the exact  microscopic Hamiltonian model but one can construct effective Hamiltonian models that encode all the dynamical properties one wishes to study. Many techniques can be followed to construct effective Hamiltonian models, most of which are based on perturbation theory and adiabatic elimination~\cite{ref:JOSA,ref:Steinbach1997,ref:Rahav2003,ref:Aniello2005,ref:Shao2017} . 
A most useful tool in these derivations is the so-called rotating wave approximation (RWA)~\cite{ref:AllenEberly}, consisting in removing some terms both on a physical ground (since they are not conserving the energy of the system) and at a mathematical level because of the appearance of fast phase factors in the interaction picture, implying negligible effects of the relevant terms in the dynamics of the system, especially when a coarse grained dynamic is to be evaluated. Suitable energy-shift terms are usually also considered as a side-effect of the elimination of the counter-rotating terms.
 
Recently there has been a growing interest in the field of non-Hermitian Hamiltonians (NHH)~\cite{ref:Bender1998,ref:DelCampo2008,ref:Rudner2009,ref:Feng2011,ref:Regensburger2012,ref:Fyod2012,ref:Gros2014,ref:Ashida2017,ref:Nakagawa2018,ref:Kawabata2019,ref:Militello2019a,ref:Militello2019b,ref:Michi2020}, most of which are the result of the elimination of some degrees of freedom, in particular cutting the relevant Hilbert space in order to focus on a specific subspace of interest. It is then interesting to understand what happens to the RWA when the Hamiltonian governing the system is non-Hermitian, since in such a case moving to the interaction picture can still produce the appearance of (fast) phase factors but also the appearance of (quick) decay factors. For a three-state system, the adiabatic elimination of a decaying state has been performed to get an effective Hamiltonian for the subsystem related to the other two states~\cite{ref:DelCampo2008}.

In this paper we consider a physical system characterized by a set of decaying levels coupled to a set of non-decaying ones, then describable by a non-Hermitian Hamiltonian model which contains complex diagonal terms. We prove that under the hypothesis of large decay rates one can neglect some terms of the Hamiltonian describing the interaction between the two subsystems, the decaying and the non-decaying ones, which we call the evanescent wave approximation (EWA). Similarly to what happens with the RWA and adiabatic elimination, some energy-shifts and effective couplings appear in the relevant effective Hamiltonian, this time constituting a non-Hermitian operator. These terms are then responsible for some effective decays.

Non-Hermitian Hamiltonians have been extensively used to describe the so-called continuous-measurement Zeno effect, where the repeated pulsed measurements on a physical system, typical of the standard quantum Zeno effect~\cite{ref:MishraSudarshan}, are replaced by some decay which play the role of a continuous monitoring of the populations of the relevant states~\cite{ref:DelCampo2008,ref:Presilla1996,ref:Home1997,ref:Schulman1998,ref:Panov1999a,ref:Panov1999b,ref:PascazioFacchi2001,ref:Militello2001,ref:PascazioFacchi2002PRL,ref:PascazioFacchi2008,ref:PascazioFacchi2009,ref:MilitelloPScr2011}. Exploiting our theory for the derivation of the effective Hamiltonian based on the EWA, we will be in a condition to interprete the occurrence of a Zeno dynamics as the negligibility of the pseudo-Lamb shifts of the system effective Hamiltonian.

%%%%%%%%%%%%%%%%%%%%%%%%%%%%%%%%%%%%%%%%%%
\section{Handling the Model}

\subsection{Non-Hermitian Hamiltonian Model}

\begin{figure}{h}
\centering
\includegraphics[width=7 cm]{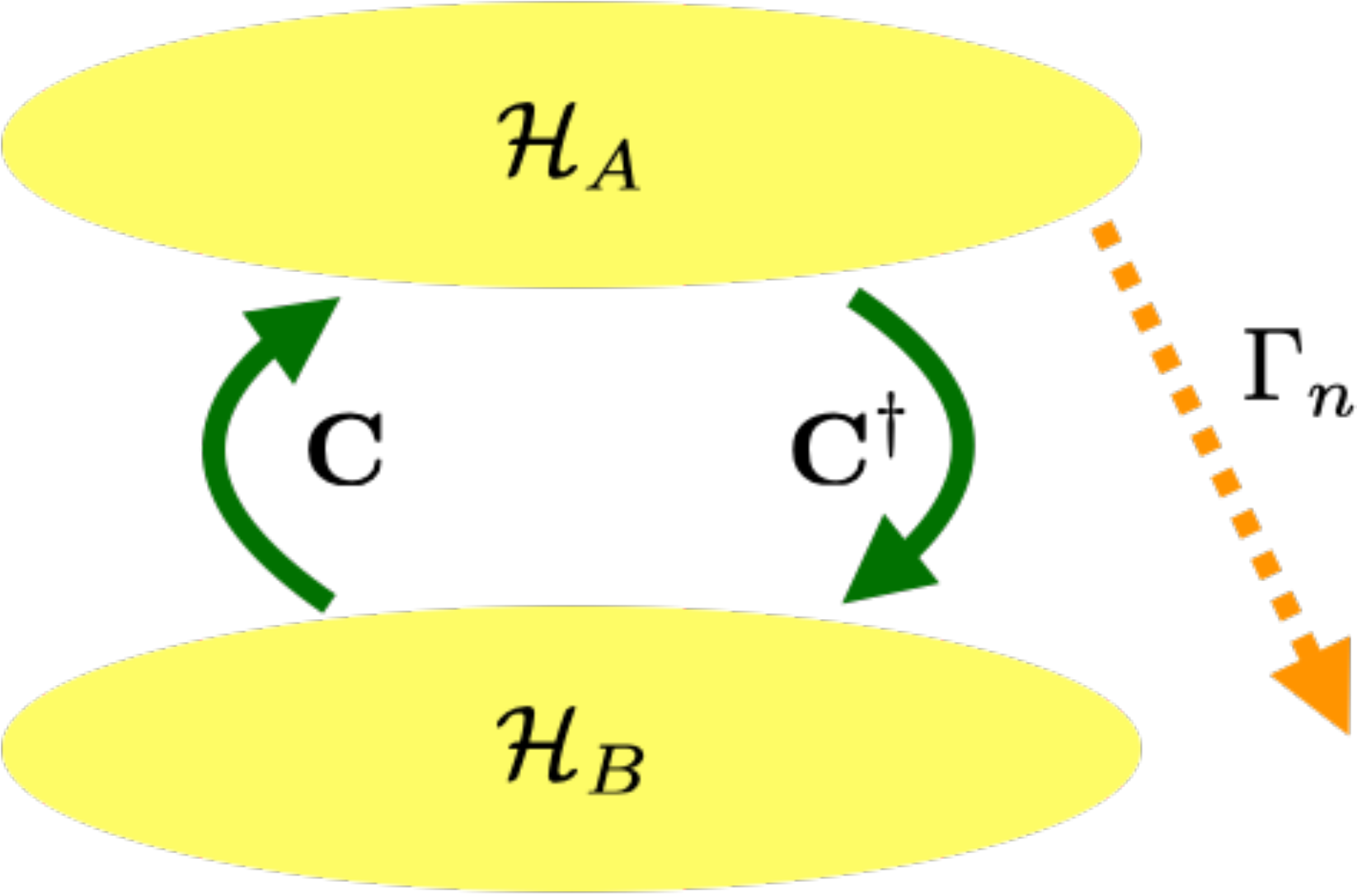} 
\caption{Pictorial representation of the system consisting of two subspaces (${\cal H}_A$ and ${\cal H}_B$, or simply $A$ and $B$) coupled via the terms $\mathbf{C}$ and $\mathbf{C}^\dag$. The subsystem $A$ is subjected to decays with rates $\Gamma_n$'s toward levels external to ${\cal H}_A \oplus {\cal H}_A$.}\label{fig:scheme}
\end{figure}

We consider a system whose relevant Hilbert space is made of two subspaces, one (${\cal H}_A$, or simply $A$) subjected to decays toward some other levels and one (${\cal H}_B$, or simply $B$) not decaying but coupled to the former one. A pictorial representation is given in Fig.~\ref{fig:scheme} while details of the derivation of the NHH describing this situation are given in Sec.~\ref{sec:DerivationNNH}.

The system Hamiltonian model admits the following block representation:
\begin{eqnarray}\label{eq:DefH0HI}
H= H_0 + H_I\,, \qquad
H_0 = 
\left(\begin{array}{cc}
\mathbf{A} & \mathbf{0} \\
\mathbf{0} & \mathbf{B}
\end{array}\right)\,, \qquad 
H_I = 
\left(\begin{array}{cc}
\mathbf{0} & \mathbf{C} \\
\mathbf{C}^\dag & \mathbf{0}
\end{array}\right)\,,
\end{eqnarray}
with $\mathbf{B}^\dag = \mathbf{B}$ while $\mathbf{A}$ is expected to be non-Hermitian ($\mathbf{A}^\dag\not=\mathbf{A}$).
For the sake of simplicity we assume $\mathbf{A}$ to have zero non-diagonal entries, while its diagonal terms are complex numbers $\mathbf{A}_{nn} = \omega_n -\ii\Gamma_n$.

The state of the system will have components both in $A$ and in $B$, which we will indicate as $\Ket{\psi_A}$ and $\Ket{\psi_B}$, so that we can write 
\begin{eqnarray}
\Ket{\psi} = \left(
\begin{array}{c}
\Ket{\psi_A} \\
\Ket{\psi_B}
\end{array}
\right) \,.
\end{eqnarray}
Accordingly, we have $\partial_t\Ket{\psi_A} = -\ii \, \mathbf{A} \Ket{\psi_A} - \ii \, \mathbf{C} \Ket{\psi_B}$ and $\partial_t\Ket{\psi_B} = -\ii \, \mathbf{B} \Ket{\psi_B} - \ii \, \mathbf{C}^\dag \Ket{\psi_A}$.

\subsection{Effective Hamiltonian: The Evanescent Wave Approximation}

Under the hypothesis of large $\Gamma_n$'s, we can derive an effective Hamiltonian which provides a closed description for the dynamics of $\Ket{\psi_B}$, i.e., an evolution not explicitly involving  $\Ket{\psi_A}$. Specifically, we require that 
\begin{eqnarray}
\Gamma \equiv \min_n\{\Gamma_n\} \gg \max_{ij} \{ |c_{ij}| \} \equiv c\,,
\end{eqnarray}
with $c_{ij} = \Bra{i} \mathbf{C}\Ket{j}$, $\Ket{i}\in {\cal H}_A$ and $\Ket{j}\in {\cal H}_B$.

Let us consider a time window $[t, t+\Delta t]$ and write down the evolution in such a time interval in an interaction picture defined by 
\begin{eqnarray}\label{eq:IPDef}
\Ket{\tilde{\psi}(s)} = \ee^{\ii H_0 (s-t)} \Ket{\psi(t)} \,,
\end{eqnarray}
so that
\begin{eqnarray}
\partial_s \Ket{\tilde{\psi}(s)} = -\ii \, \tilde{H}_I(s) \Ket{\tilde{\psi}(s)} \,,
\end{eqnarray}
with the transformed interaction term
\begin{eqnarray}
\tilde{H}_I (s) = \ee^{\ii H_0 (s-t)} H_I \ee^{-\ii H_0 (s-t)} \,.
\end{eqnarray}
Though one could expect having the exponential of $H_0^\dag$ on the right of the operator, the Hamiltonian in the new picture is obtained according with the need of inserting the identity $\ee^{-\ii H_0 (s-t)} \ee^{\ii H_0 (s-t)}$ between $H_I$ and the state, in the interaction-picture Schr\"odinger equation.

The second-order approximated state at time $t+\Delta t$ is given by:
\begin{eqnarray}
\nonumber
\Ket{\tilde{\psi}(t+\Delta t)} &=& \Ket{\tilde{\psi}(t)} - \ii \int_t^{t+\Delta t}\tilde{H}_I(\xi)\Ket{\tilde{\psi}(t)} \mathrm{d}\xi - \int_t^{t+\Delta t}  \mathrm{d}\xi \int_t^\xi \tilde{H}_I (\xi) \tilde{H}_I (\eta) \Ket{\tilde{\psi}(\eta)} \mathrm{d}\eta \\
&\approx& \Ket{\tilde{\psi}(t)} - \ii \int_t^{t+\Delta t}\tilde{H}_I(\xi)\Ket{\tilde{\psi}(t)} \mathrm{d}\xi - \int_t^{t+\Delta t}  \mathrm{d}\xi \int_t^\xi \tilde{H}_I (\xi) \tilde{H}_I (\eta) \Ket{\tilde{\psi}(t)} \mathrm{d}\eta
\end{eqnarray}
from which we can obtain an approximated expression for the time derivative of the state in the interaction picture:
\begin{eqnarray}
\partial_t \Ket{\tilde{\psi}(t)} 
&\approx& -\ii \, \tilde{H}_I(t)\Ket{\tilde{\psi}(t)}  - \int_t^{t+\Delta t} \tilde{H}_I (t+\Delta t) \tilde{H}_I (\eta) \Ket{\tilde{\psi}(t)} \mathrm{d}\eta \,,
\end{eqnarray}
which means having $\partial_t \Ket{\tilde{\psi}(t)} = \tilde{H}_{\mathrm{eff}}(t) \Ket{\tilde{\psi}(t)}$ with the following effective Hamiltonian:
\begin{eqnarray}\label{eq:Heff_operatorial}
\tilde{H}_{\mathrm{eff}}(t) = \tilde{H}_I(t)  - \ii \int_t^{t+\Delta t} \tilde{H}_I (t+\Delta t) \tilde{H}_I (\eta)  \mathrm{d}\eta \,.
\end{eqnarray}
This Hamiltonian admits the following block representation:
\begin{eqnarray}
\tilde{H}_{\mathrm{eff}}(t) = 
\left(\begin{array}{cc}
-\ii \int_t^{t+\Delta t} \mathrm{d}\eta \tilde{\mathbf{C}}_\uparrow(t+\Delta t) \tilde{\mathbf{C}}_\downarrow(\eta) & \tilde{\mathbf{C}}_\uparrow(t+\Delta t) \\
\tilde{\mathbf{C}}_\downarrow(t+\Delta t) & -\ii  \int_t^{t+\Delta t} \mathrm{d}\eta \tilde{\mathbf{C}}_\downarrow(t+\Delta t) \tilde{\mathbf{C}}_\uparrow(\eta)
\end{array}\right)\,,
\end{eqnarray}
with 
\begin{eqnarray}
\tilde{\mathbf{C}}_\uparrow(s) &=& \ee^{\ii H_0 (s-t)} \, \mathbf{C} \, \ee^{-\ii H_0 (s-t)} \,, \\
\tilde{\mathbf{C}}_\downarrow(s) &=& \ee^{\ii H_0 (s-t)} \, \mathbf{C}^\dag \, \ee^{-\ii H_0 (s-t)} \,.
\end{eqnarray}
Note that, because of $H_0$ being non Hermitian, it is $\tilde{\mathbf{C}}_\uparrow(s) \not= \tilde{\mathbf{C}}^\dag_\downarrow(s)$. Moreover, since we are considering an effective Hamiltonian in the interaction picture defined by \eqref{eq:IPDef}, the diagonal terms do not involve the matrix blocks $\mathbf{A}$ and $\mathbf{B}$ of $H_0$. Finally, as a result of the second-order perturbation treatment, we get the diagonal blocks \, --- which are \lq\lq non-Hermitian operatorial dressings\rq\rq, including pseudo-energy shifts and effective couplings --- \,  from the second term in the right-hand side of \eqref{eq:Heff_operatorial}.

As shown in Sec.~\ref{sec:neglectpsia}, it turns out that $\|\Ket{\tilde{\psi}_A (t)}\|$ is small at every time. Therefore we can neglect the off-diagonal operator $\tilde{\mathbf{C}}_\downarrow (t+\Delta t)$, since it is small itself (containing only vanishing exponentials) and acting on $\Ket{\tilde{\psi}_A (t)}$.  This approximation, based on the assumption $\ee^{-\Gamma_n \Delta t} \ll 1$ is the essence of the EWA. Such a treatment is still valid even if $\|\Ket{\psi_A (0)}\|$ is not negligible, since $\|\Ket{\psi_A (t)}\|$ rapidly vanishes anyway, as shown in Sec.~\ref{sec:neglectpsia}. Moreover, even if $\|\Ket{\psi_A (t)}\|$ were simply bound (not necessarily small) we would have anyway that the terms coming from the action of $\tilde{\mathbf{C}}_\downarrow (t+\Delta t)$ on $\Ket{\psi_A(t)}$ are small.

Coming back to the Schr\"odinger picture, we eventually obtain the following effective Hamiltonian:
\begin{eqnarray}
H_{\mathrm{eff}} = 
\left(\begin{array}{cc}
\mathbf{A} -\ii \, \mathbf{D}_A & \mathbf{C} \\
\mathbf{0} & \mathbf{B}  -\ii \, \mathbf{D}_B
\end{array}\right)\,,
\end{eqnarray}
with 
\begin{eqnarray}
%\nonumber
\label{eq:DADef}
\mathbf{D}_A &=&  \ee^{-\ii H_0 \Delta t} \, \tilde{\mathbf{C}}_\uparrow (t+\Delta t) \left[ \int_t^{t+\Delta t} \mathrm{d}\eta \tilde{\mathbf{C}}_\downarrow (\eta) \right] \ee^{\ii H_0 \Delta t} \,, \\
\label{eq:DBDef}
\mathbf{D}_B &=& \ee^{-\ii H_0 \Delta t} \, \tilde{\mathbf{C}}_\downarrow (t+\Delta t) \left[ \int_t^{t+\Delta t} \mathrm{d}\eta \tilde{\mathbf{C}}_\uparrow (\eta) \right] \ee^{\ii H_0 \Delta t} \,.
\end{eqnarray}

The operator $\mathbf{D}_B$ contains many evanescent wave terms identified by the presence of decaying exponentials $\ee^{-\Gamma_n \Delta t}$. Therefore, assuming that $\Delta t$ is large enough to consider all such terms vanishing (the same assumption that allowed us to neglect $\mathbf{C}_\downarrow$ in the interaction picture), we get the EWA expression for $\mathbf{D}_B$:
\begin{eqnarray}\label{eq:DBexpr}
\mathbf{D}_B &\approx& \mathbf{D}_B^{EWA} = \sum_{m m'} \sum_n \, \frac{c_{nm}^* c_{nm'}}{\Gamma_n + \ii (\omega_n - \omega_{m'})} \, \KetBra{m}{m'} \,.
\end{eqnarray}
This block-matrix contains both diagonal and non-diagonal terms, thus describing both shifts of pseudo-energies (i.e., energy shifts and decay rates) and effective couplings between states of $B$. Eq.~\eqref{eq:DBexpr} is applicable also in the presence of bare-energy degeneracies, because the negligibility of the terms discarded from \eqref{eq:DBDef} and the non vanishing of the denominator in \eqref{eq:DBexpr} are both guaranteed by the presence of the decay rates.

The validity of our approach relies on the EWA valid for $\Gamma_n \gg |c_{ij}|$, nevertheless similar results can be obtained also with the RWA, based on $|\omega_n-\omega_m| \gg |c_{ij}|$, which allows for neglecting the rapidly-oscillating coupling terms (in this case, not only $\mathbf{C}_\downarrow$, but also $\mathbf{C}_\uparrow$). Moreover, in some cases, the two approximations can cooperate. Anyway, in this work we are focusing on the EWA.

\subsection{Quantum Zeno Effect}

Under the hypotheses justifying the derivation of our effective Hamiltonian, we have that the equation governing the evolution of $\Ket{\psi_B}$ is closed, in the sense that it does not involve $\Ket{\psi_A}$. In fact, it turns out that
\begin{eqnarray}
\partial_t \Ket{\psi_B(t)} = -\ii \, \mathbf{H}_B^{EWA}  \Ket{\psi_B(t)} \,,
\end{eqnarray}
and we introduce $\mathbf{H}_B^{EWA} = \mathbf{B} - \ii \mathbf{D}_B^{EWA}$ as the EWA Hamiltonian for the subspace coupled to the decaying one.

It is worth noting that for high values of $\Gamma_n$'s %--- which is our case --- 
the operator $\mathbf{D}_B^{EWA}$ is pretty small, though not necessarily negligible. Anyway, the higher the values of the $\Gamma_n$'s, the smaller the entries of the operator $\mathbf{D}_B^{EWA}$, as it immediately follows from \eqref{eq:DBexpr}. In this case, the dynamics of the B-component of the global state $\Ket{\psi_B}$ turns out to be governed by the sole operator $\mathbf{B}$, meaning that such component evolves as if the interaction with the other levels (the subspace $A$) were not present, i.e., as if $\mathbf{C}=\mathbf{0}$ in \eqref{eq:DefH0HI}. This is the signature of a continuous measurement quantum Zeno effect, where the decays play the role of measurements: having higher $\Gamma_n$'s is the continuous counterpart of having more frequent pulsed measurements on the states of the subspace $A$, which neutralizes the interaction between $A$ and $B$.

These occurrences have been already predicted in other works~\cite{ref:DelCampo2008,ref:Presilla1996,ref:Home1997,ref:Schulman1998,ref:Panov1999a,ref:Panov1999b,ref:PascazioFacchi2001,ref:Militello2001,ref:PascazioFacchi2002PRL,ref:PascazioFacchi2008,ref:PascazioFacchi2009,ref:MilitelloPScr2011} and recently analyzed with the exploitation of a perturbative treatment~\cite{ref:MiliNapoPLA2020}. The EWA approach anyway puts the phenomenon in a different light, tracing back the Hilbert space partitioning due to high diagonal terms to an approximation similar to the RWA, in some aspects, but differing from the RWA because of the presence of decaying exponentials instead of oscillating ones.

%%%%%%%%%%%%%%%%%%%%%%%%%%%%%%%%%%%%%%%%%%

\section{Numerical analysis}

We now apply the theory developed in the previous section initially focusing on two simple systems: a three-state and a four-state ones. In both cases the subsystem $B$ is characterized by two states, $\Ket{1}$ and $\Ket{2}$, corresponding to the eigenvalue $0$ and $\epsilon$, respectively, whereas the subsystem $A$ is either one-dimensional, involving only the state $\Ket{3}$ characterized by the decay rate $\Gamma_3$ and bare energy $\omega_3$, or two-dimensional involving also the state $\Ket{4}$ corresponding to $\Gamma_4$ and $\omega_4$. The two states of $B$ are coupled, which implies the presence of nonzero off-diagonal terms $\Bra{1} \mathbf{B}\Ket{2} = \Bra{2} \mathbf{B}\Ket{1}^* \equiv g$.

\begin{figure}[h]
\centering
\begin{tabular}{ccc}
\subfigure[]{\includegraphics[width=7 cm]{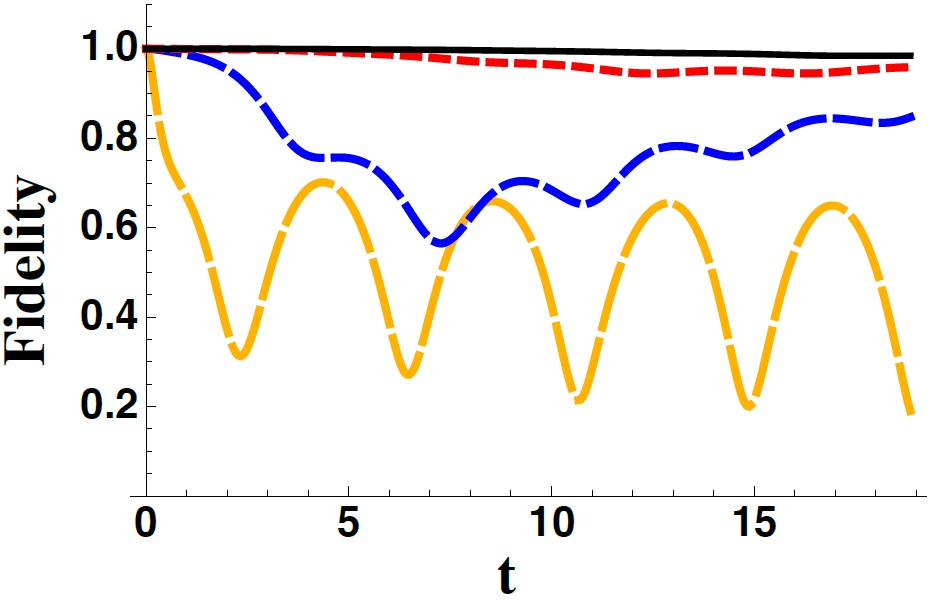}}  & \qquad& \subfigure[]{\includegraphics[width=7 cm]{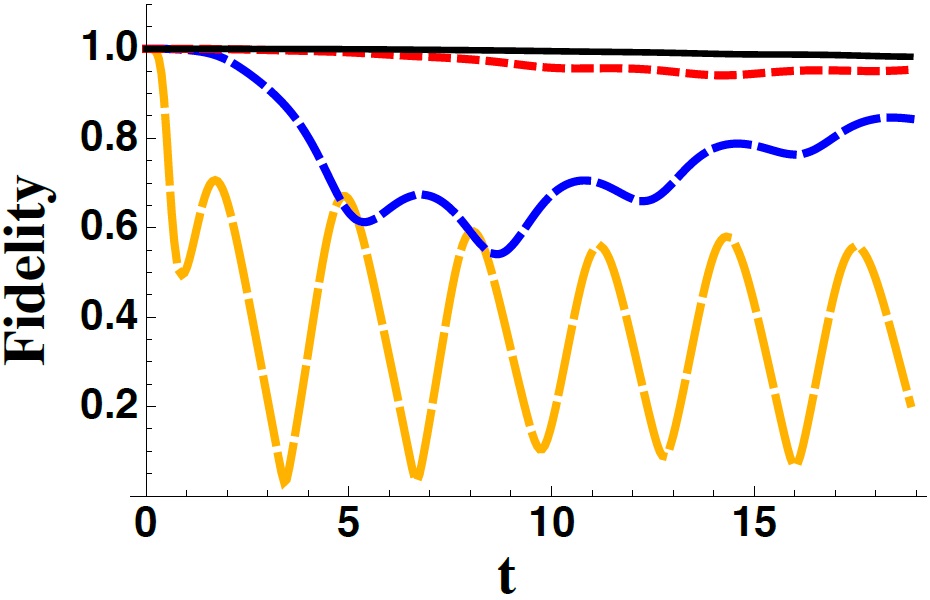}} \\
\subfigure[]{\includegraphics[width=7 cm]{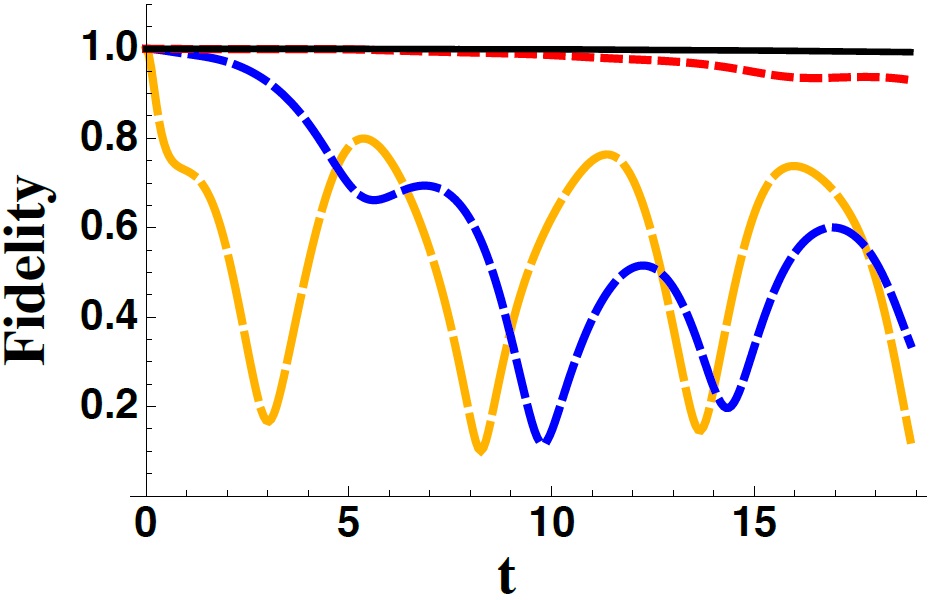}}  & \qquad & \subfigure[]{\includegraphics[width=7 cm]{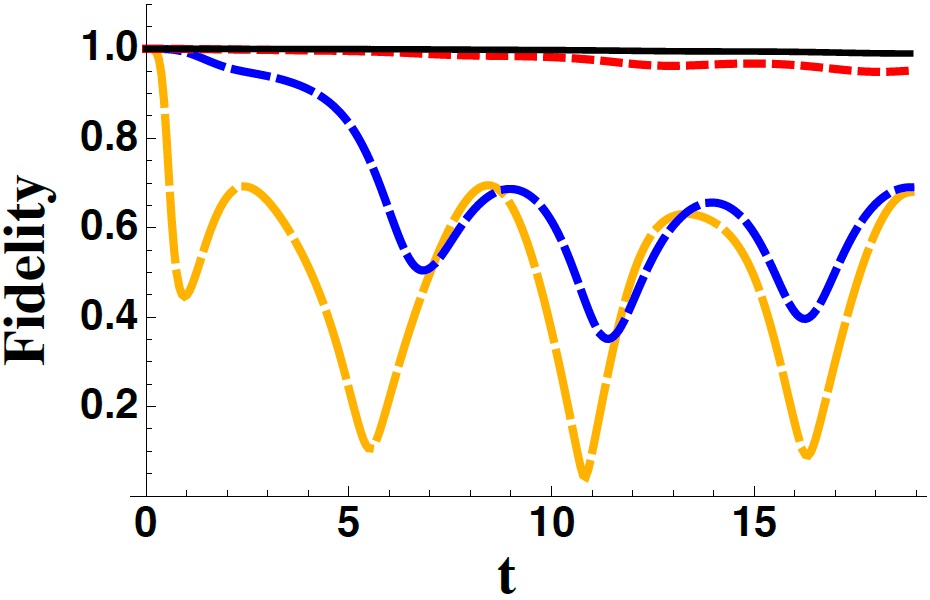}} 
\end{tabular}
\caption{ ${\cal F}_{EWA}$ for $\Gamma_3/\epsilon = 0.1$ (orange long-dashed line), $\Gamma_3/\epsilon = 1$ (blue dashed line),  $\Gamma_3/\epsilon = 3$ (red dotted line),  $\Gamma_3/\epsilon = 5$ (black solid line), assuming $\Ket{\psi} = p_A \Ket{3} + \sqrt{1-p_A} (\cos\theta \Ket{2}+\sin\theta \Ket{1})$ as the initial states, and considering different values of the parameters: $g/\epsilon = 0.5$, $p_A=0$, $\theta=0$ (a); \,  $g/\epsilon = 0.5$, $p_A=0$, $\theta=\pi/4$ (b); \, $g/\epsilon = 0.25$, $p_A=0$, $\theta=0$ (c); \,  $g/\epsilon = 0.25$, $p_A=0.25$, $\theta=\pi/4$ (d). In all plots we have considered $c_{31}=c_{32}=0.5\, \epsilon$. Time is in units of $\epsilon^{-1}$.
}\label{fig2}
\end{figure}

\begin{figure}[h]
\centering
\begin{tabular}{ccc}
\subfigure[]{\includegraphics[width=7 cm]{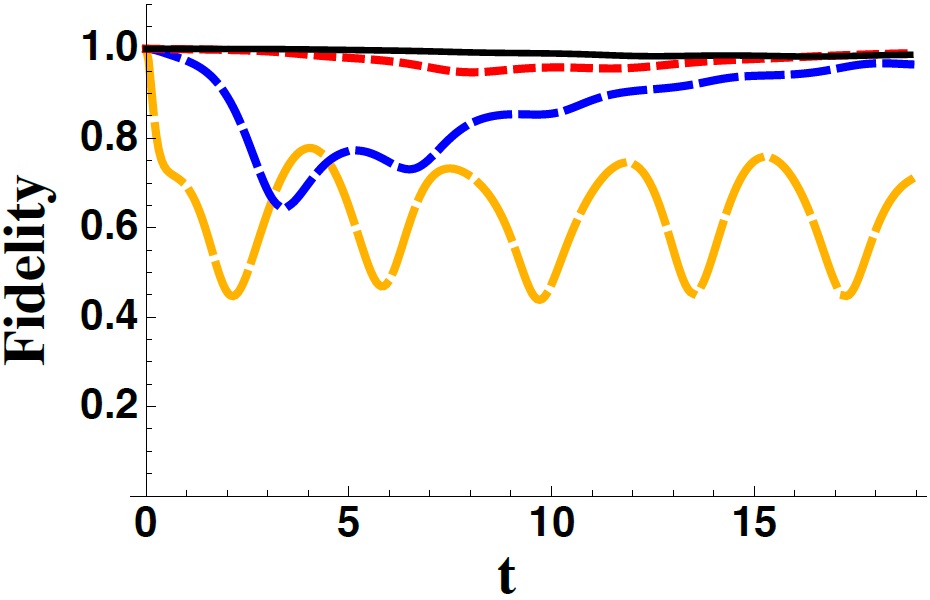}} & \qquad& \subfigure[]{\includegraphics[width=7 cm]{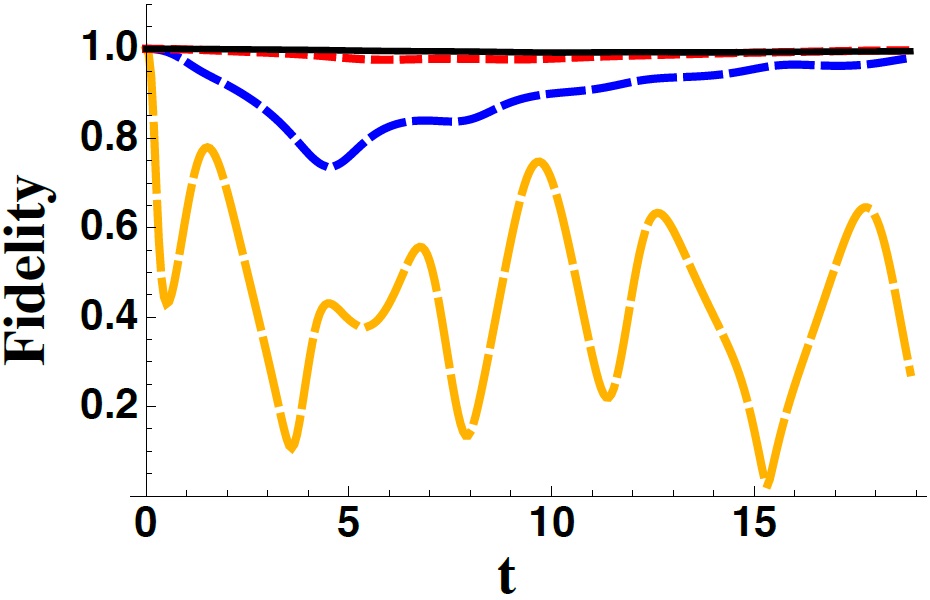}} 
\end{tabular}
\caption{ ${\cal F}_{EWA}$ for $\Gamma_3/\epsilon = 0.1$ (orange long-dashed line), $\Gamma_3/\epsilon = 1$ (blue dashed line),  $\Gamma_3/\epsilon = 3$ (red dotted line),  $\Gamma_3/\epsilon = 5$ (black solid line), assuming $\Ket{\psi} = p_A \Ket{3} + \sqrt{1-p_A} (\cos\theta \Ket{2}+\sin\theta \Ket{1})$ as the initial states, and considering different values of the parameters: $g/\epsilon = 0.5$, $p_A=0$, $\theta=0$ (a);  $g/\epsilon = 0.5$, $p_A=0.1$, $\theta=\pi/3$ (b). In all plots we have considered $c_{31}=c_{32}=0.5\, \epsilon$, $c_{41}=c_{42}=0.4 \, \epsilon$ and $\Gamma_4/\Gamma_3=1.2$. Time is in units of $\epsilon^{-1}$. 
}\label{fig3}
\end{figure}

In Figs.~\ref{fig2} and \ref{fig3} we show the fidelity between the exact and the effective dynamics, generated by $H$ and $H_\mathrm{eff}$ respectively, corresponding to different parameter values as well as different initial conditions. In particular, we focus on the fidelity between the $B$-components of the states, hence evaluating the following quantity:
\begin{equation}
{\cal F}_{EWA} (t) = 
\frac{\left| \Bra{\psi(0)} \ee^{\ii \, H^\dag t} \, \hat{\Pi}_B \,  \ee^{-\ii \, \mathbf{H}_B^{EWA} t} \Ket{\psi(0)} \right| }{ 
|\Bra{\psi(0)} \ee^{\ii \, H^\dag t} \, \hat{\Pi}_B \, \ee^{-\ii \, H t} \Ket{\psi(0)}|^{1/2} \times
|\Bra{\psi(0)} \ee^{\ii \, (\mathbf{H}_B^{EWA})^\dag t} \, \hat{\Pi}_B \, \ee^{-\ii  \mathbf{H}_B^{EWA} t}  \Ket{\psi(0)}|^{1/2}
} \,,
\end{equation}
where $\hat{\Pi}_B$ is the projector operator onto the subspace $B$ and we have considered the normalization factors (the denominator) for the two projected states. Such normalization is necessary because the two states can have norms smaller than unity (both because we project on $B$ and because the dynamics itself contains decays), and, on the other hand, what we are interested in is the coincidence between the two evolutions in $B$, irrespectively of the fact that the relevant components of the states have norms smaller than unity.

Because of the hypothesis $\Gamma \gg c$, the effective Hamiltonian derived through our treatment is not supposed to be valid for small values of $\Gamma$. Nevertheless, we have used our effective model also for small values of the $\Gamma_n$'s in order to check the failure of our treatment in the small-damping limit and to observe at which point it becomes valid.
In Figs.~\ref{fig2}  we show the behaviour of ${\cal F}_{EWA}$ for the three-state system in different situations, and corresponding to four different values of the decay rate, that is $\Gamma_3/\epsilon = 0.1$, $\Gamma_3/\epsilon = 1$,  $\Gamma_3/\epsilon = 3$ and $\Gamma_3/\epsilon = 5$.
In all situations (different couplings and different initial conditions), we observe that for the higher considered value of the decay rate, $\Gamma_3/\epsilon = 5$ which also corresponds to $\Gamma_3/c_{31} = \Gamma_3/c_{32} = 10$, the fidelity is kept very close to unity at every time, confirming the validity of the description given by our effective Hamiltonian. 
The four plates refer to different situations: in (b) and (d), superpositions of the states $\Ket{1}$ and $\Ket{2}$ are considered, whereas in (a) and (c) the initial state projected on $B$ is always $\Ket{2}$. In (c) and (d) a different value of the coupling constant $g$ with respect to (a) and (b) is considered. Finally, in (d) we assumed an initial condition with a non vanishing $\Ket{\psi_A}$ component.
It is the case to emphasize that the effectiveness of the hamiltonian derived following the procedure discussed in the previous section is very satisfactory even at moderately-high values of $\Gamma_3$. The fidelity indeed is very close to unity at any time even for $\Gamma_3/\epsilon=3$, becoming essentially unity for $\Gamma_3/\epsilon=5$. We have moreover checked that the figures coming from other initial conditions or choosing different values of $c_{31}$ and $c_{32}$ as well as $g$, show all the same qualitative behaviour, provided the general hypotheses of our theory for getting an effective Hamiltonian are satisfied (mainly $\Gamma\gg c$). 

In Fig.~\ref{fig3} we plot the fidelity between the complete and effective dynamics obtained assuming that the subspace $A$ is two-dimensional. In order to make effective the presence of the fourth level, we consider nonzero values of the off-diagonal entries $c_{41}$ and $c_{42}$, which implies a coupling between the subspace $B$ and the new state $\Ket{4}$ which decays with rate $\Gamma_4$.  Also in this case the fidelity is very close to unity even for moderately-high values of $\Gamma_3$ and $\Gamma_4$.
In fact, increasing the dimension of the subspace $A$ does not change the qualitative behaviour of ${\cal F}_{EWA}$.  
We thus can conclude that, as far as the evolution of the subsystem $B$ is considered, we can adopt the EWA Hamiltonian even in the presence of high but not very high values of the decay rates.

\begin{figure}[h]
\centering
\begin{tabular}{ccc}
\subfigure[]{\includegraphics[width=7 cm]{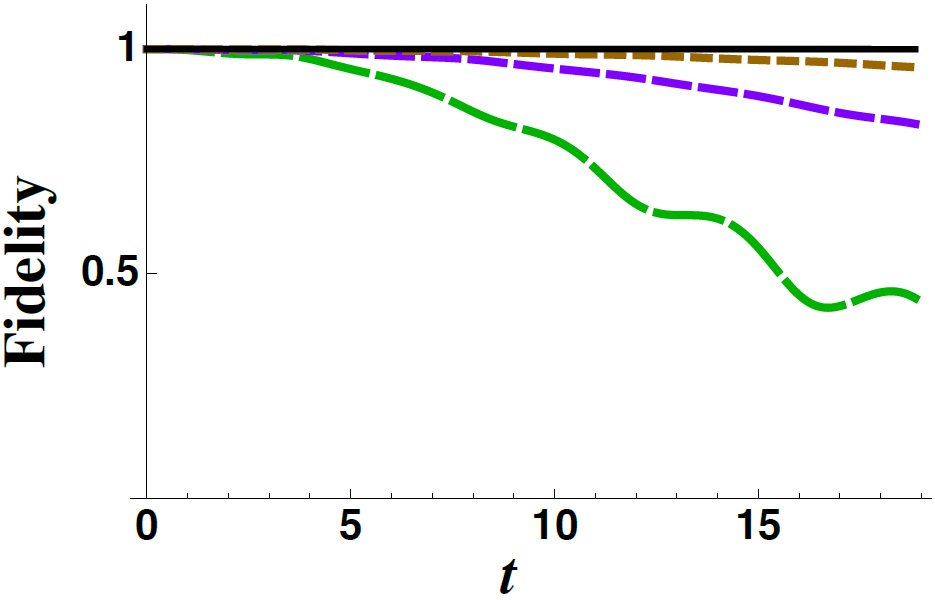}} & \qquad& \subfigure[]{\includegraphics[width=7 cm]{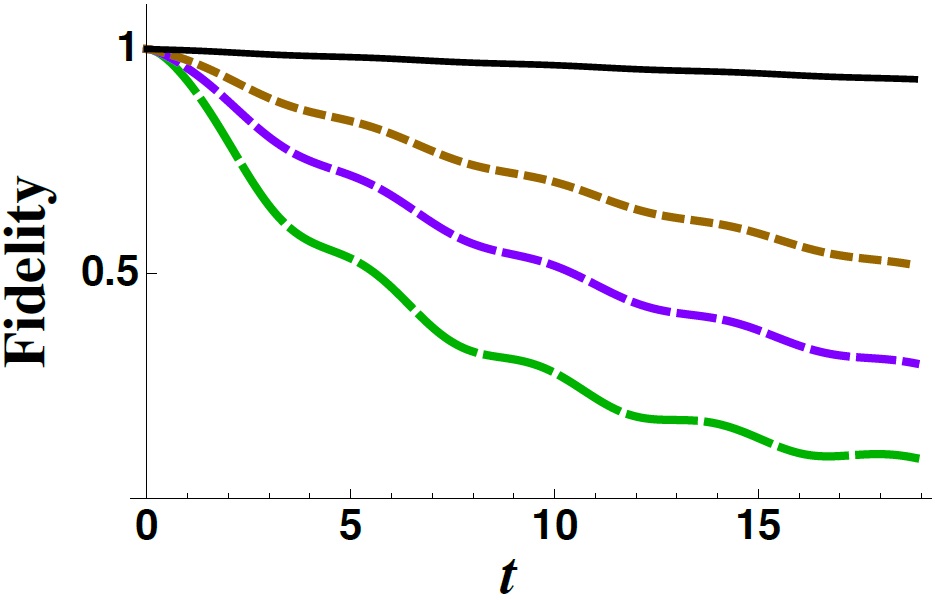}} 
\end{tabular}
\caption{Fidelities ${\cal F}_{ZN}$ (a) and ${\cal F}_Z$ (b) as functions of time (in units of $\epsilon^{-1}$) for different values of the decay rates: $\Gamma_3/\epsilon = 2$ (green long-dashed line), $\Gamma_3/\epsilon = 5$ (purple dashed line), $\Gamma_3/\epsilon = 10$ (brown dotted line), $\Gamma_3/\epsilon = 100$ (black solid line). The other parameters have the following values: $g = 0.5 \, \epsilon$, $c_{31}=c_{32}=0.5\, \epsilon$, $c_{41}=c_{42}=0$, $\Gamma_4=0$ and $\Ket{\psi(0)} = \Ket{2}$.
}\label{fig4}
\end{figure}

As previously discussed, equation (\ref{eq:DBexpr}) puts into evidence the fact that, making the decay rates larger and larger, the dynamics of the subspace $B$ turns out to be effectively governed by the sole operator $\mathbf{B}$, thus evolving as if the coupling with $A$ was absent ($\mathbf{C}=\mathbf{0}$). This behaviour can be interpreted as the occurrence of a Zeno dynamics. 
Such prediction is corroborated by the fidelities evaluated in Fig.~\ref{fig4} (and many others not reported in the present work). In this case we compare the complete dynamics generated by $H$, projected into the subspace $B$, with the unperturbed dynamics generated by $\mathbf{B}$, starting with a state in $B$. Indeed,  our point is that thanks to the validity of the effective description given by $\mathbf{H}_B^{EWA}$ and the fact that $\mathbf{H}_B^{EWA} \rightarrow \mathbf{B}$ for high values of $\Gamma_n$'s, we expect that the fidelity between the complete dynamics and that generated by $\mathbf{B}$ converges to unity.

In connection with the Zeno effect, it is useful to consider both the fidelities obtained normalizing or not the state projected on $B$ after the evolution governed by $H$. Indeed, on the one hand it is important to understand whether the projection on $B$ of the evolved states via $H$ is equivalent to the projection of the state evolving via $\mathbf{B}$, but on the other hand it is interesting to know whether the dynamics induced by $H$ in the subspace $B$ is equivalent to that induced by $\mathbf{B}$ up to some global decay factor, which we can remove through a renormalization (see Ref.~\cite{ref:MiliNapoPLA2020} for an extensive discussion). 
Therefore, it is useful to define the fidelity for the non-normalized state,
\begin{eqnarray}
{\cal F}_{Z} (t) &=& 
\left| \Bra{\psi(0)} \ee^{\ii \, H^\dag t} \, \hat{\Pi}_B \, \ee^{-\ii \, \mathbf{B} t} \Ket{\psi(0)} \right|  \,, 
\end{eqnarray}
and the fidelity for the normalized state,
\begin{eqnarray}
{\cal F}_{ZN} (t) &=& 
\frac{\left| \Bra{\psi(0)} \ee^{\ii \, H^\dag t} \, \hat{\Pi}_B \, \ee^{-\ii \, \mathbf{B} t} \Ket{\psi(0)} \right| }{ 
|\Bra{\psi(0)} \ee^{\ii \, H^\dag t} \, \hat{\Pi}_B \, \ee^{-\ii \, H t} \Ket{\psi(0)}|^{1/2} \times
|\Bra{\psi(0)} \ee^{\ii \, \mathbf{B} t} \hat{\Pi}_B \ee^{-\ii  \mathbf{B} t}  \Ket{\psi(0)}|^{1/2}
} \,.
\end{eqnarray}
The second normalizing factor in the denominator is useful only if the system has an initial state with a nonzero component in $A$.

It is well visible that for very large values of the decay rates, the unperturbed dynamics ($\mathbf{C}=\mathbf{0}$) and the complete one essentially coincide. Nevertheless, it is worth stressing that, although both the occurrence of  a Zeno dynamics and the validity of the effective description based on EWA are related to high values of the $\Gamma$'s, it is well visible that the latter occurs for smaller values of the decay rates (moderately high value), while the former requires very high values.

%\unskip
%\subsubsection{Subsubsection}

 In all figures (\ref{fig2}, \ref{fig3} and \ref{fig4}) we have considered $\omega_3=\omega_4=0$, for the sake of simplicity. Nevertheless, we have made other plots with nonzero values for such frequencies obtaining the same qualitative results.

%%%%%%%%%%%%%%%%%%%%%%%%%%%%%%%%%%%%%%%%%%
%\section{Materials and Methods}
\section{Methods}

\subsection{Derivation of the Non-Hermitian Hamiltonian}\label{sec:DerivationNNH}

Here we present the derivation of the NHH for a system with a subspace undergoing decays, following the same procedure of Ref.~\cite{ref:MiliNapoPLA2020}.
The fact that the subspace $A$ is subjected to decays toward states not belonging to $B$ is expressed by the presence of some couplings between the states of ${\cal H}_A$ and the states of another subspace ${\cal H}_G$ (or simply $G$), that we assume to have energies lower than those of $A$; we are also assuming zero temperature, so that there are only decays from $A$ toward $G$, while there is no thermal pumping from $G$ toward $A$. Finally, we assume no coherent coupling between $G$ and the other two subspaces ($A$ and $B$).
The relevant master equation may be written as~\cite{ref:Gardiner,ref:Petru}
\begin{equation}\label{eq:NStateSystemME}
\dot\rho = -\ii [H_S, \rho^S] + \sum_{ij} \gamma_{ij} \left( \hat{X}_{ij} \rho^S \hat{X}_{ij}^\dag - \frac{1}{2} \{ \hat{X}_{ij}^\dag \hat{X}_{ij}, \rho^S\} \right) \,,
\end{equation}
where $H_S$ is the Hermitian Hamiltonian of the system (whose total Hilbert space is ${\cal H}_S = {\cal H}_A \oplus {\cal H}_B \oplus {\cal H}_G$), $\hat{X}_{ij}$ are suitable jump operators connecting states of $A$ with states of $G$, $\gamma_{ij}$ being the relevant decay rates.  
Only the terms with $\hat{X}_{ij} = \hat{\Pi}_{G} \hat{X}_{ij} \hat{\Pi}_{A}$ ($\hat\Pi_k$, $k=A,G$ are the projectors onto the relevant subspaces) have non vanishing $\gamma_{ij}$, due to the structure of the $\hat{X}_{ij}$ and to the zero-temperature hypothesis. On the contrary,  the terms with $\hat{X}_{ij} = \hat{\Pi}_{A} \hat{X}_{ij} \hat{\Pi}_{G}$ are absent.

All considered, projecting on ${\cal H}_A \oplus {\cal H}_B$, we obtain a closed equation for the density operator $\rho^{AB} = (\hat{\Pi}_A + \hat\Pi_B) \, \rho \, (\hat{\Pi}_A + \hat\Pi_B)$, which is:
\begin{equation}\label{eq:PseudoLiouville}
\partial_t \rho^{AB} = -\ii (H_0 \rho^{AB} - \rho^{AB} H_0^\dag)\,, 
\end{equation}
with
\begin{equation}\label{eq:NStateSystemRedME}
H_0 =  (\hat{\Pi}_A + \hat\Pi_B) \hat{H}_S  (\hat{\Pi}_A + \hat\Pi_B)  - \ii \sum_{ij} \frac{\gamma_{ij}}{2} \hat{X}_{ij}^\dag \hat{X}_{ij} \,.
\end{equation}

If we now add an interaction involving the states of $A$ and $B$ and described by $H_I$ and rename $\rho^{AB}$ as $\rho$, we get 
\begin{equation}\label{eq:HDefNHH}
H = H_0 + H_I \,,
\end{equation}
and 
\begin{equation}\label{eq:PseudoLiouville}
\partial_t \rho = -\ii (H \rho - \rho H^\dag)\,,
\end{equation}
which admits the solution
\begin{equation}\label{eq:Evolution}
 \rho(t) = \ee^{-\ii H t} \rho(0) \ee^{\ii H^\dag t}\,.
\end{equation}

As a consequence, the dynamics when the system starts in a pure state $\rho = \KetBra{\psi_0}{\psi_0}$ can be evaluated as the non-unitary dynamics governed by the equation $\partial_t \Ket{\psi} = -\ii H \Ket{\psi}$, with the initial condition $\Ket{\psi(0)} = \Ket{\psi_0}$.
It is worth emphasizing that we haven't made any approximation to derive the non-Hermitian Hamiltonian and that using it to describe the dynamics of the subspace $A+B$ is perfectly equivalent to exploiting the master equation. This happens because at zero temperature the subspace $A$ undergoes dissipation and decoherence,  which can be well described by adding suitable imaginary contributions to the diagonal entries of the Hamiltonian. On the other hand, focusing on the dynamics of  $A+B+G$ would make impossible to obtain an effective Hamiltonian, since the dynamics of $G$ is characterized by incoherently receiving population from $A$, which cannot be obtained by any Hamiltonian, even a non-Hermitian one.

\subsection{Smallness of $\| \Ket{\psi_A} \|$}\label{sec:neglectpsia}

Here we prove that the assumption $\|\Ket{\tilde{\psi}_A(t)}\|\approx 0$ at every time, provided it is $\| \Ket{\psi_A(0)} \| \approx 0$, is consistent with the EWA. Since, on the basis of \eqref{eq:IPDef} we have $\Ket{\tilde{\psi}(t)}=\Ket{\psi(t)}$, we will focus on proving the smallness of $\|\Ket{\psi_A(t)}\|$.  

Let us start by considering that both $\|\Ket{\psi_A(t)}\|\le 1$ and  $\|\Ket{\psi_A(t)}\| \le 1$ hold for all $t$, as a consequence of the facts that $ \mathrm{tr}(\KetBra{\psi(0)}{\psi(0)})=1$ and $\partial_t \mathrm{tr}(\KetBra{\psi}{\psi}) = -\ii \, \mathrm{tr}(H\KetBra{\psi}{\psi} - \KetBra{\psi}{\psi} H^\dag) = - 2 \sum_n \Gamma_n |\BraKet{a_n}{\psi}|^2 \le 0$, where $\Gamma_n$'s are the real parts of the eigenvalues of $H$ and $\Ket{a_n}$'s are the relevant eigenstates.

Now, for the coefficients $\BraKet{a_n}{\psi_A}$ we obtain the following set of differential equations:
\begin{eqnarray}
\partial_t \BraKet{a_n}{\psi_A(t)} = (- \Gamma'_{n} - \ii\omega'_{n} ) \, \BraKet{a_n}{\psi_A(t)} - \ii \, \sum_{m} \Bra{a_n} \mathbf{C}' \Ket{m} \, \BraKet{m}{\psi_B(t)} \,.
\end{eqnarray}

The formal solution of the $n$-th of such equations is:
\begin{eqnarray}
\nonumber
\BraKet{a_n}{\psi_A(t)} &=& \exp{[(- \Gamma_{n} - \ii\omega_{n} ) t]} \, \Bigg[ \BraKet{a_n}{\psi_A(0)} \\
\nonumber
&-& \ii \, \int_{0}^t  \mathrm{d} s \exp{[(\Gamma_{n} +\ii\omega_{n} ) s]} \sum_{m} \Bra{a_n} \mathbf{C}\Ket{m} \, \BraKet{m}{\psi_B(s)} \Bigg] \\
\nonumber
&\le& \big| \exp{[(- \Gamma_{n} - \ii\omega_{n} ) t]} \big| \, \times \, \Bigg[ \Big| \BraKet{a_n}{\psi_A(0)} \Big| 
\\
\nonumber
&+&  \int_{0}^t  \mathrm{d} s  \times \, \bigg| \exp{[(\Gamma_{n} +\ii\omega_{n} ) s]} \sum_{m} \Bra{a'_n} \mathbf{C} \Ket{m} \, \BraKet{m}{\psi_B(s)} \bigg| \Bigg] \\
%\nonumber
&\le&  \exp(- \Gamma_{n} t) \, \left| \BraKet{a_n}{\psi_A(0)}  \right|  +  \frac{1 - \exp{ (- \Gamma_{n} t) }}{|\Gamma_{n} + \ii\omega_{n}|} \sum_{m} |\Bra{a_n} \mathbf{C} \Ket{m} | \,,
\end{eqnarray}
where we have used that $|\BraKet{m}{\psi_B(t)}| \le 1$. 

Assuming $ \BraKet{a_n}{\psi_A(0)} \sim o(c/\Gamma)$ we have $ \BraKet{a_n}{\psi_A(t)} \sim o(c/\Gamma)$ at any time. Therefore, $\|\Ket{\tilde{\psi}_A(t)}\| = \|\Ket{\psi_A(t)}\| \sim o(c/\Gamma)$.
It is worth noting that even in the case where $\BraKet{a_n}{\psi_A(0)}$ is not small, the exponential factor $\exp(-\Gamma_n t)$ kills its contribution to $\BraKet{a_n}{\psi_A(t)}$ after a very short time, therefore making essentially irrelevant the modulus of $\BraKet{a_n}{\psi_A(t)}$ at any further time.

%%%%%%%%%%%%%%%%%%%%%%%%%%%%%%%%%%%%%%%%%%
\section{Conclusions}

The rotating wave approximation is one of the most popular way to treat Hamiltonian models difficult to deal with, since the presence of the so called counter-rotating terms can make quite hard the resolution of the dynamics. This approximation is based on the fact that such counter-rotating terms acquire, in the interaction picture, very fast time-dependent phase factors, whose average effect on the system time evolution can be neglected, at some extent. When the Hamiltonian governing the system is non-Hermitian, the passage to the interaction picture produces in general the appearance of phase factors and of decays. Some terms can then become negligible because of the rapid oscillations or because of the vanishing of their moduli. In connection with this second occurrence, we have introduced the evanescent wave approximation, consisting in neglecting the terms acquiring decay factors ($\tilde{\mathbf{C}}_\downarrow \approx \mathbf{0}$) and introducing a sort of dressing and pseudo-Lamb shifts ($\mathbf{D}_A$ and $\mathbf{D}_B$). Differently from the standard situations, this dressing in our case is non-Hermitian too, resulting in the appearance of effective decays in the subspace which originally does not undergo any direct decay (subsystem $B$).  All these facts considered, an effective Hamiltonian is introduced, allowing for evaluating the time evolution in the non-decaying subspace with a close equation not involving the components of the state which belong to the decaying subsystem (i.e., a closed equation for $\Ket{\psi_B}$ is obtained). This is essentially an adiabatic elimination of the decaying levels (subspace $A$) allowed by the fact that the decaying states are very low populated during all the evolution.

We have observed that the EWA is a very good approximation, allowing for predictions very close to those obtained from the original Hamiltonian, even in the presence of moderately-high decay rates. Moreover, thanks to the way we can write the evolution in the non-decaying subspace, we can predict a quantum Zeno effect as the diminishing of the dressing obtained for increasing decay rates. In fact, higher values of the $\Gamma_n$'s imply $\mathbf{D}_B$ to become more and more negligible.

% The following MDPI journals use author-date citation: Arts, Econometrics, Economies, Genealogy, Humanities, IJFS, JRFM, Laws, Religions, Risks, Social Sciences. For those journals, please follow the formatting guidelines on http://www.mdpi.com/authors/references
% To cite two works by the same author: \citeauthor{ref-journal-1a} (\citeyear{ref-journal-1a}, \citeyear{ref-journal-1b}). This produces: Whittaker (1967, 1975)
% To cite two works by the same author with specific pages: \citeauthor{ref-journal-3a} (\citeyear{ref-journal-3a}, p. 328; \citeyear{ref-journal-3b}, p.475). This produces: Wong (1999, p. 328; 2000, p. 475)

%%%%%%%%%%%%%%%%%%%%%%%%%%%%%%%%%%%%%%%%%%
%% optional
%\sampleavailability{Samples of the compounds ...... are available from the authors.}

%% for journal Sci
%\reviewreports{\\
%Reviewer 1 comments and authors’ response\\
%Reviewer 2 comments and authors’ response\\
%Reviewer 3 comments and authors’ response
%}

%%%%%%%%%%%%%%%%%%%%%%%%%%%%%%%%%%%%%%%%%%
\end{document}